%% file: main.tex
\documentclass[sigconf]{acmart}

\AtBeginDocument{%
  }

\copyrightyear{2026}
\acmYear{2026}
\setcopyright{cc}
\setcctype{by}
\acmConference[ICPC '26]{34th IEEE/ACM International Conference on Program Comprehension}{April 12--13, 2026}{Rio de Janeiro, Brazil}
\acmBooktitle{34th IEEE/ACM International Conference on Program Comprehension (ICPC '26), April 12--13, 2026, Rio de Janeiro, Brazil}
\acmDOI{10.1145/3794763.3798175}
\acmISBN{979-8-4007-2482-4/2026/04}

\usepackage{mdframed}
\usepackage{multirow}
\usepackage{hyperref}

\begin{document}

\title{Agile Story-Point Estimation: Is RAG a Better Way to Go?}

\author{Lamyea Maha}
\affiliation{%
  \institution{University of Saskatchewan}
  \city{Saskatoon}
  \state{SK}
  \country{Canada}
}
\email{lamyea.maha@usask.ca}

\author{Tajmilur Rahman}
\affiliation{%
  \institution{Gannon University}
  \city{Erie}
  \state{PA}
  \country{USA}
}
\email{rahman007@gannon.edu}

\author{Chanchal Roy}
\affiliation{%
  \institution{University of Saskatchewan}
  \city{Saskatoon}
  \state{SK}
  \country{Canada}
}
\email{chanchal.roy@usask.ca}

\ccsdesc[500]{Software and its engineering~Requirements analysis}

\begin{abstract}
The sprint-based iterative approach in the Agile software development
method allows continuous feedback and adaptation.
One of the crucial Agile software development activities is the sprint-planning
session where developers estimate the effort required to complete
tasks through a consensus-based estimation technique such as Planning Poker. 
In the Agile software development method, a common unit of measuring development effort is \textbf{Story Point (SP)} which is assigned to tasks to understand the complexity and development time needed to complete them. 
Despite the benefits of this process, it is an extremely time-consuming manual process. 
To mitigate this issue, in this study, we investigated if this manual process can be automated using Retrieval Augmented Generation (RAG) which comprises a ``Retriever'' and a ``Generator''.
We applied two embedding models - \textit{bge-large-en-v1.5}, and Sentence-Transformers' \textit{all-mpnet-base-v2} on 23 open-source software projects of varying sizes and examined four key aspects: 1) how retrieval hyper-parameters influence the performance, 2) whether estimation accuracy differs across different sizes of the projects, 3) whether embedding model choice affects accuracy, and 4) how the RAG-based approach compares to the existing baselines. 
Although the RAG-based approach outperformed the baseline models in several occasions, our results did not exhibit statistically significant differences in performance across the projects or across the embedding models. 
This highlights the need for further studies and refinement of the RAG, and model adaptation strategies for better accuracy in automatically estimating user stories. 
\end{abstract}

\keywords{RAG, LLM, Agile, Effort Estimation, Software Engineering}
\maketitle

\input{01_introduction}
\input{02_background}
\input{03_methods}
\input{04_evaluation}
\input{05_results}

\input{06_threats}
\balance{}
\input{07_conclusions}

\input{08_acknowledgement}


\bibliographystyle{ACM-Reference-Format}
\bibliography{main}

\end{document}

%% file: 01_introduction.tex
\section{Introduction}
Many organizations have been using agile methodology for their software development processes for a long time~\cite{dybaa2008empirical}. The agile development setting allows developers to work by sprints (iterations of two-weeks cycle) during which team members complete a specific set of tasks~\cite{kumar2012impact} from product backlog items (tasks).
This iterative approach allows developers to receive customer feedback frequently and test features early~\cite{chan2009acceptance}. Based on the feedback from the customer, they create new tasks for improvements that go to the backlog and then plan for the next sprint. This discussion session is typically known as the sprint planning session that takes place right before each sprint begins. Due to the progressive nature of agile development processes, it is important that the development team knows how much effort is required to complete the tasks in the upcoming sprints which will help to utilize available resources accurately~\cite{usman2014effort}.
Planning and estimation in agile processes is different compared to traditional (non-agile) development processes~\cite{cohn2005agile}. 
One of the most studied and widely used effort estimation techniques during the sprint planning session in the agile development process is ``Planning Poker''~\cite{usman2014effort} which allows developers to assess product backlogs and agree on the required effort to complete each task~\cite{mahnivc2012using}. 
The metric used for effort estimation is called \textbf{Story Point (SP)}. In this study, our focus is to automatically estimate SPs to reduce the time and effort developers need during the sprint planning session.
Story points are measured in numbers that indicate the overall effort in terms of hours and the complexity of the task~\cite{trendowicz2014software}. 
This helps the team plan for the next sprint~\cite{tawosi2022investigating}.

Planning Poker session involves people from different levels of experience, such as junior developers, senior developers, team leaders, software testers, scrum masters, and project managers. 
This diversity gives the team better estimations of story points as it reduces over-optimism that might be exerted by the more experienced developers.~\cite{molokken2004group} 
However, this process may introduce biases during effort estimation~\cite{jorgensen2004review}. 
For example, factors such as the influence of senior team members, or the presence of dominant personalities during the meeting can affect the story point estimations by the junior members which can lead to issues like cost overruns, idle developer time, inefficient use of resources, or project failure~\cite{porru2016estimating}. 
Moreover, Planning Poker can be time-consuming too. According to Cohn~\cite{cohn2005agile}, generally, it takes between 5 and 10 minutes to estimate each task in the product backlog. 
This means, for a project like Linux Kernel, Planning Poker session would take about 6 months~\cite{alhamed2021playing}. To address the challenges of the manual sprint planning approach, there is a growing interest in using machine learning for estimating story points~\cite{ritu2023software}. 
Automated systems can make use of historical data to offer consistent, unbiased, and efficient estimates of tasks more efficiently~\cite{porru2016estimating}. 
%
Previous studies have primarily focused on either using different features of a task to estimate the story points or learning from the dataset of pre-estimated tasks s to produce an estimation~\cite{porru2016estimating, choetkiertikul2018deep}. 
In another study~\cite{fu2022gpt2sp} although the authors have tried to make the story point estimation process interpretable, they have only worked on the titles of the tasks. 
Moreover, a replication study~\cite{tawosi2024agile} has shown that there is a bug in that study which affected the reported results. 
In this study, we explored how well a Retrieval-Augmented Generation (RAG) based story point estimation approach would work. To the best of our knowledge, there are no previous studies which explored RAG based story point estimation. In this approach, there are two components, the retriever and the generator. First, given a new task, the retriever uses embedding models to generate embeddings and hence retrieves similar tasks from the embeddings of the training dataset of a particular project. We then used Llama-3.2-3B-Instruct as the generator component to estimate story points based on the retrieved tasks. So, in our study, we have tried to explicitly retrieve and represent concrete examples that include both titles and descriptions of the tasks—just like developers reviewing similar tasks that they have already estimated before voting during a planning poker session. 
Our approach would also ensure interpretability and transparency in effort estimation using story points since the developers can see the retrieved similar tasks based on which the generator has based its answers on mimicking the widely used planning poker session.

However, before such an automated approach can be considered a reliable alternative to established human-based estimation practices such as Planning Poker, it is necessary to rigorously evaluate its effectiveness. Planning Poker incorporates the collective knowledge, negotiation, and experience of developers, whereas a RAG-based method removes the human element entirely and relies solely on retrieved context and model inference. This raises the question of whether such a system can produce estimates of comparable quality in real-world settings. To investigate this, we formulated the following research questions and a set of hypotheses to conduct a systematic empirical study across multiple open-source projects and project sizes. Our research questions are as follows:\\

\textbf{RQ1 - How do the retrieval parameters (top\_k and temperature) influence performance, and how do the optimal parameter settings differ across project size groups?}\\
Retrieval-augmented generation relies on the quality and quantity of contextual information retrieved from historical project artifacts. In our study, we intend to find if there are any optimal values for top\_k (number of retrieved similar instances) and temperature (generation determinism) for which RAG based story point estimation gives the highest accuracy. However, the effectiveness of retrieval parameters such as top\_k and temperature may vary depending on the scale and diversity of available project data. Projects of different sizes offer varying levels of redundancy and contextual richness, meaning that a parameter setting that works well for a small project may not be optimal for a large one. Therefore, examining how these parameters influence performance across project size categories is essential to establish practical guidance for configuring RAG systems in real-world software development environments. We also intend to use the findings from this RQ to answer the rest of the RQs below.

\textbf{RQ2 - Does the effectiveness of the RAG approach for story point estimation differ across small, mid-sized, and large software projects?}\\
As discussed in RQ1, software projects vary widely in size, and the availability of historical tasks can influence how well retrieval-based techniques perform. In smaller projects, the historical repository may not provide enough relevant examples for effective retrieval, while very large projects may introduce noise and redundancy, making it harder to retrieve the most contextually useful tasks. Since RAG mainly depends on retrieving similar past tasks as context, its performance may be sensitive to the scale of the project on which it is applied. Understanding whether RAG behaves differently across small, mid-sized, and large projects is therefore essential for determining its practical applicability in real-world development environments. If performance degrades for certain project sizes, practitioners would need guidelines or adaptations before adopting the approach. On the other hand, if RAG performs consistently across varying scales, it would indicate robustness and wider deployment potential. With this research question, we intend to find out if selecting the best suited parameters according to the project size makes significant improvement in story point estimation.

\textbf{RQ3 - Does the choice of embedding models significantly affect the estimation accuracy?}\\
Different embedding models capture semantic information in different ways, and these differences can influence how effectively similar historical tasks are retrieved during the RAG story point estimation process. Since the retrieval step directly affects the quality of the generated story point estimation, the choice of embedding model is not a trivial implementation detail but a critical design decision. In this study, we experiment with two embedding models - Beijing Academy of Artificial Intelligence’s (BAAI) \textit{bge-large-en-v1.5} and sentence-transformers’ (SBERT) \textit{all-mpnet-base-v2}. In the rest of the paper we refer to Beijing Academy of Artificial Intelligence’s (BAAI) and sentence-transformers’ (SBERT) \textit{all-mpnet-base-v2} embedding model as BAAI and SBERT respectively.
By examining whether these two embedding models produce embeddings that lead to significantly different estimation accuracies, we can determine whether choosing the right embedding model could provide practical benefits in effort estimation tasks. This insight would guide both researchers and practitioners in selecting appropriate models that balance performance with computational cost, ultimately improving the reliability and applicability of RAG-based estimation systems in real-world software development settings.

\textbf{RQ4 - How well does our approach perform compared to the existing methods?}\\
To assess the effectiveness of our RAG approach to estimating SPs, we need to conduct a comparative analysis against established approaches. The four baseline methods with which we compared our results are: a classical machine learning method~\cite{porru2016estimating} (TF-IDF), a deep learning technique~\cite{choetkiertikul2018deep} (Deep-SE), and two clustering-based methods~\cite{tawosi2022investigating} (LHC-SE and LHCtc-SE ).

We have made the replication package of this study available on a public repository\footnote{https://doi.org/10.6084/m9.figshare.30842972}.

\subsection{Hypotheses}
\label{hypothesis}
To empirically evaluate the effectiveness and behavior of our RAG based story point estimation approach, we formulated statistical hypotheses corresponding to the research questions: RQ2, RQ3, and RQ4. 
These hypotheses guide our experimental design and allow us to assess whether the observed differences in performance are statistically meaningful.

\subsubsection{Hypothesis corresponding to RQ2}
Our second research question investigates whether RAG based story point estimation differ across small, mid-sized, and large projects. 
In this regard, we hypothesized that our RAG based approach will show significantly different story-point estimation performance across different-sized projects. 
We considered Mean Absolute Errors (MAE) as the measure of model's performance. 
Hence, our null Hypothesis was stated as:

\textit{$H'_0$: There is no significant difference between the MAEs obtained from the RAG based models across different size-groups of the projects under study.}

and, the alternative Hypothesis was stated as:

\textit{$H'_1$: There is a significant difference between the MAEs obtained from the RAG based models across different size-groups of the projects under study.}

This hypothesis allows us to examine whether the effectiveness of the RAG approach is sensitive to the scale of the project repository.

\subsubsection{Hypothesis corresponding to RQ3}
Our third research question assesses whether the embedding models, BAAI and SBERT, have a significant impact on the resulting MAEs.
In this regard, we hypothesized that the accuracy of the story point estimation by SBERT and BAAI would differ.

Hence, our null Hypothesis was stated as:

\textit{$H''_0$: There is no significant difference between the accuracy of BAAI and SBERT in predicting the story points for the given projects.}

and, the alternative Hypothesis was stated as:

\textit{$H''_1$: The embedding models result in significantly different accuracies.}

This hypothesis allows us to determine whether the choice of embedding model plays a substantial role on the effectiveness of story point estimation.

\subsubsection{Hypothesis corresponding to RQ4}
Our fourth and the final research question compares the RAG-based models with four baseline models: Deep-SE, LHC-SE, LHCtc-SE, and TF-IDF.
In this regard, we hypothesized that the performance of the RAG-based approach will outperform the baseline models.

Hence, our null Hypothesis was stated as:

\textit{$H'''_0$: The MAEs of the RAG-based models are not significantly lower than the MAE of the existing baseline methods.}

and, the alternative Hypothesis was stated as:

\textit{$H'''_1$: The MAEs of the RAG-based approach are significantly lower than the MAEs of the baseline models.}

This hypothesis helps to determine whether RAG provides a performance advantage over state-of-the-art approaches and whether its improvements are statistically significant.


The rest of the paper is structured as follows: section~\ref{sec:related_work} describes the works that are related to our study, section~\ref{sec:methodology} describes the methodology that we designed, section~\ref{sec:exp_design_results} describes the experimental design and the results that we obtained, section~\ref{sec:threats_to_validity} describes the threats to validity for our study and how we addressed them, and finally section~\ref{sec:conclusion} concludes the paper with future directions.

%% file: 02_background.tex
\section{Background and Related Works}
\label{sec:related_work}
\subsection{Story Points}
A ``Story Point'' (SP) is a unit to measure the effort required to complete a task in the Agile software development method.
SP is a compound measure of relative complexity, risk, and size of a task, encouraging teams to think more abstractly about estimated effort required~\cite{ijais12-450574}. 
During sprint planning, developers assign story points based on task scope, uncertainty, and technical challenges. 
For simplicity, in this paper we have used `tasks' indicating any of the Agile work items such as, Epics, Stories, Tasks, Sub-tasks, and Bugs.

In Planning Poker, developers use numbered cards (typically following a Fibonacci sequence~\cite{porru2016estimating}) to propose story point estimates for a task. 
After a brief discussion, they reveal their estimates simultaneously. 
Discrepancies in estimates are discussed further, helping the team refine their understanding of the task and reach a consensus. 
This process not only promotes shared ownership of the estimates but also helps identify potential risks or misunderstandings early. 
Despite revealing cards simultaneously, and including developers from different expertise, the SP can still be affected by human biases~\cite{jorgensen2004review}.

Story points also indicate how difficult or complex a task is for the development team in terms of resources and uncertainty. 
They are not objective measures and hence, story points for a particular task may be different for different development team~\cite{porru2016estimating}. 
Tasks with higher numbers indicate that they require more effort to complete compared to tasks with smaller numbers. 
\subsection{Story Point Based Effort Estimation Techniques}
Several SP-based effort estimation techniques are commonly used in agile development processes. 
Among these, \textit{Planning Poker}, \textit{Use Case Points Estimation}, and \textit{Expert Judgment} are the most popular~\cite{usman2014effort}. 
Although these techniques share similar activities and goals, choosing the right one depends on the specific project needs. 
To determine the most suitable approach, we relied on the recommendations from a survey conducted on 10 experienced software developers. 
Based on their recommendations, we selected Planning Poker as the preferred technique.

Planning Poker is a collaborative Agile estimation technique where team members, such as developers, testers, and managers, individually estimate tasks to build shared understanding~\cite{mahnivc2012using}. 
It leverages collective input to reduce over-optimism and produce more realistic estimates.
Despite its benefits, it can be biased by optimism, anchoring, or dominant voices, leading to cost overruns, resource mismanagement, or project failure~\cite{porru2016estimating}. Its effectiveness also depends on team dynamics and clear communication.

\subsection{Story Point Estimation Approaches}
Several studies used machine learning approaches to estimate story points.
One of the earliest works on automatic estimation of agile effort was introduced by Abrahamsson et al.~\cite{abrahamsson2011predicting}. 
The authors used factors such as, number of characters present in the tasks and priority, along with 15 most frequently used keywords. 
They used these factors to train 3 machine learning models.

Porru et al.~\cite{porru2016estimating} used TF-IDF (Term-Frequency \& Inverse Document Frequency transformation) to extract features from task title and description. 
They also found that attributes such as `type of task' and the `components of the task' are useful for story point estimation. 
The authors experimented with classical machine learning models, including, Support Vector Machine (SVM), K-Nearest Neighbor, Decision Tree, and Naive Bayes to identify the best performing model. 
Among these, SVM was reported to perform better. This study is referred to as the ~\textit{TF-IDF} approach in this study. 

Deep learning based story point estimation models have also been explored. 
Choetkiertikul et al.~\cite{choetkiertikul2018deep} proposed \textit{Deep-SE} which uses a combination of Long Short-Term Memory (LSTM) with Recurrent Highway Net (RHWN) to estimate story points. 
They trained the model with the titles and descriptions of the tasks. 
LSTM finds the vector representation of the description and title and RHWN is further used to learn the deep representation of these vectors. 
The final vector is then fed to the linear regression model which estimates the story point for the task. 
They also published the dataset of tasks that they worked on, which contains 23,313 tasks up until August 8th, 2016, collected from 16 open source projects. 
In this paper from here on we refer this study as the ``\textit{Deep-SE} study''.

Fu et al.~\cite{fu2022gpt2sp} used a GPT-2-based model for story point estimation on the dataset by Choetkiertikul et al.~\cite{choetkiertikul2018deep}, although a replication study~\cite{tawosi2024agile} found a bug that made their results unreliable, so we did not compare with their data and approach. 
Our study addresses three \textit{Deep-SE} limitations: reliance on project-specific vocabulary, one-directional LSTM processing that misses semantic context, and lack of interpretability in its predictions.

Tawosi et al.~\cite{tawosi2022investigating} proposed \textit{LHC-SE}, a story point estimation method using LDA for topic modeling and hierarchical clustering. Each task is represented by topics, clustered, and estimated using the cluster's median. They extended this to \textit{LHC$_\text{TC}$-SE} by incorporating task type and component information to improve accuracy.

In another study~\cite{phan2022heterogeneous}, the authors designed an algorithm to produce a graph from an issue and the authors used heterogeneous graph neural networks to estimate story points. They used the dataset provided by Choetkiertikul et al.~\cite{choetkiertikul2018deep}.

In another study of Tawosi et al.~\cite{tawosi2023search}, they applied Search-Based Software Engineering to improve story point estimation with LLMs such as GPT-4 by optimizing a small set of example tasks. However, the same example set is reused for all inputs, without dynamically selecting examples per task.

Our study aims to explore story point estimation using a Retrieval-Augmented Generation (RAG) approach. 
Tawosi et al.~\cite{tawosi2022investigating} evaluated four methods—\textit{Deep-SE}, \textit{TF-IDF}, \textit{LHC-SE}, and \textit{LHC$_\text{TC}$-SE}—on their custom dataset. 
We use their reported results as benchmarks for comparison with our proposed approach.

%% file: 03_methods.tex
\section{Methodology}
\label{sec:methodology}

\begin{figure*}[ht!]
\includegraphics[width=\textwidth]{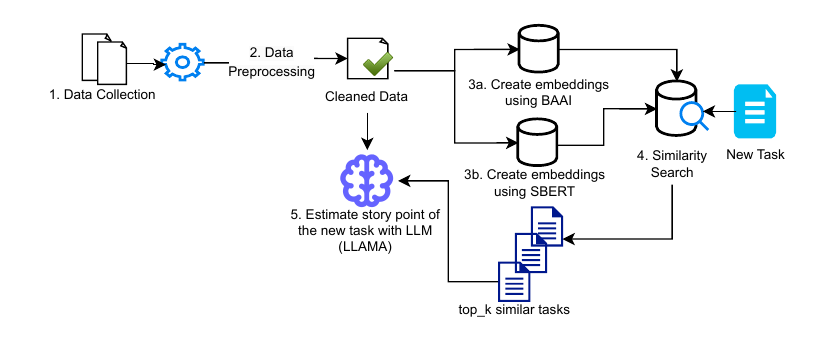}
\caption{Methodology}
\label{fig:methodology}
\end{figure*}

Our methodology begins with data collection. We utilized the Tawosi dataset comprising tasks from open-source projects that follow the Agile development process. 
The details of the data collection process are elaborated in Section~\ref{sec:data}. 
Following a series of pre-processing steps, we split the dataset into a training set (80\%) and a testing set (20\%) for model evaluation. Since the data is time-ordered, a single split allows training data to strictly precede test data chronologically. This design reflects real-world story point estimation, where only historical tasks are available at estimation time. Applying k-fold cross-validation would mix tasks across time, potentially allowing the model to train on or retrieve information from future tasks. To generate vector embeddings from the training set, we employed two embedding models - SBERT and BAAI. For each new task in the test set, we created its corresponding embeddings and used it to identify the most similar tasks from the embeddings of the training set.
Figure~\ref{fig:methodology} shows our methodology at a glance.

%

\subsection{Data}
\label{sec:data}
We chose to work with the dataset provided by Tawosi et al.~\cite{tawosi2022deep}, as it is more robust and comprehensive compared to other available datasets. Using this dataset, Tawosi et al.~\cite{tawosi2022investigating} experimented with four different approaches—\textit{TF-IDF, Deep-SE, LHC-SE, and LHCtc-SE}, and compared their performance. The dataset from Tawosi et al.~\cite{tawosi2022deep} includes 31,960 tasks from 26 open-source projects, significantly surpassing the dataset published by Choetkiertikul et al.~\cite{choetkiertikul2018deep}, which contains 23,313 tasks. 
Tawosi's dataset is more recent and hence contains more tasks. 

Moreover, to enhance the dataset's reliability, stricter filtering criteria have been applied that are recommended by Porru et al.~\cite{porru2016estimating} that not only enhance reliability but also make it more suitable for our analysis. 
The details of the recommended filtering steps are as follows:
\begin{enumerate}
    \item Tasks with updated story points or changes to fields like \textit{Description} and \textit{Summary} after SP assignment are excluded, as they may indicate initial confusion and lead to model instability.
    \item Only tasks marked as ``addressed'' based on the Status and Resolution fields are included, as unaddressed tasks may confuse estimation models~\cite{porru2016estimating}.
    \item Planning Poker cards are numbered similar to the Fibonacci series as: 0, 0.5, 1, 2, 3, 5, 8, 13, ... $\infty$. The last filtering step is to remove the issues which have story points outside Fibonacci series. 
\end{enumerate}

Although initially Tawosi's dataset had 26 projects according to the paper, 3 projects were removed from their replication package since they were removed from the public domain. 
Therefore, currently, the replication package has 12 repositories with 23 projects. 
We have used all of those 23 projects in our study.

To pre-process the data, we checked for tasks that had no description, or that did not have any SP assigned.
These tasks were originally collected from the widely popular project management system ``\textbf{Atlassian Jira}''. 
%

\subsection{The Retriever Module of RAG}
Given a new task, the retriever module is responsible for identifying the most similar tasks from the dataset. 
To achieve this, we generated embeddings of the task's textual information, such as its title and description using both BAAI and SBERT. 
Unlike standard BERT, which generates contextual embeddings for individual tokens, SBERT produces embeddings for the entire sentence, making it more efficient for directly computing semantic similarity between sentences~\cite{reimers2019sentence}. While sentence-level embeddings can also be derived by standard `BERT' model aggregating token embeddings~\cite{reimers2019sentence}, we opted for `SBERT' due to its simplicity and suitability for this task. 
We used one of the sentence transformers \textbf{\textit{all-mpnet-base-v2}}\footnote{\url{https://huggingface.co/sentence-transformers/all-mpnet-base-v2}} published in Huggingface as it is reported to give better accuracy in different semantic similarity tasks~\cite{Pretrain33:online}. Similarly, when choosing the other embedding model, we chose the BAAI's \textit{bge-large-en-v1.5} \footnote{\url{https://huggingface.co/BAAI/bge-large-en-v1.5}} model, prioritizing accuracy.



For each new task for which we wanted to estimate the SP, we performed a similarity search on the embeddings generated by the two models. 
Semantic similarity was quantified using cosine similarity, a widely used metric that measures the angle between two vectors in the embedding space. 
This approach allowed us to retrieve tasks that are most semantically similar to the given task based on its title and description. 
For each query task, we retrieved the top\_k most similar tasks where k is the number of similar tasks retrieved. We experimented with k=2,3, and 4 similar tasks.  

The following is an example of estimating story points for a task from the project `\textit{Spring XD}' in our test dataset.\\

\noindent
\textbf{Example of retrieved similar tasks:}
\begin{mdframed}
\textbf{Task Title}
Update spring-data-hadoop version to 2.1.0.RC1
\textbf{Task Description}
Update spring-data-hadoop version to 2.1.0.RC1. This also includes updating the following:  - adding hadoop26 (Apache Hadoop 2.6.0) as distro - adding hdp22 (Hortonworks HDP 2.2) as distro - set default distro to hadoop26 - update cdh5 to version 5.3.0 - remove older distros - hadoop24, hdp21\\
\textbf{List of Similar Tasks:}\\
    \textbf{1. Task Title}\\ 
    Update `spring data hadoop' dependency and add new Hadoop distros.\\
    \textbf{Task Description} 
    Update to Spring for Apache Hadoop 2.0 RC3 Add support for new hadoop distros:  - Pivotal HD 2.0 (phd20) - Hortonworks HDP 2.1 (hdp21) - Cloudera CDH5 (cdh5).\\
    \textbf{Story Point:} 8\\
    \textbf{2. Task Title}\\ 
    Update to spring-data-hadoop 1.0.1.RELEASE\\
    \textbf{Task Description} 
    This might mean we should adjust our hadoopDistro options to the ones supported in the new release - hadoop12 (default), cdh4, hdp13, phd1 and hadoop21\\
    \textbf{Story Point:} 3\\
    \textbf{3. Task Title}\\ 
    Upgrade to spring-data-hadoop 1.0.1.RC1\\
    \textbf{Task Description} 
    spring-data-hadoop 1.0.1.RC1 provides flavors for commonly used Hadoop distros/versions and we should make use of that.\\
    \textbf{Story Point:} 1\\
\end{mdframed}
\subsection{The Generator Module of RAG}
\label{sec:generator}
For the generator module, we selected LlamA-3.2-3B-Instruct as the generator model because it provides a practical balance between performance and computational efficiency. The model is instruction-tuned and performs well on reasoning and summarization tasks while remaining lightweight. We fed the retrieved tasks from the retriever module to the generator module. To estimate story point for a given task, we carried out prompt engineering to find out the prompt that can generate the best output from the model.
We tried to provide instructions to the model using different prompts with clear specifications, and detailed instructions to the model. 
We observed the performance of different versions of the prompt to find the best performing one. We tried prompts on data from different software projects in our dataset. We noticed that Llama-3.2-3B-Instruct needed explicit instruction to produce the story points in the required format. The system prompt is a static prompt which was the same for all projects. The user prompt is parameterized with three parameters. The formatted\_similar\_tasks which are retrieved by the retriever module, contain top\_k similar tasks in a pre-defined format. The other two parameters are the title of the new tasks and the description of the new task. We provide the full system prompt and the user prompt that we used for reproducibility. 
\newline\newline
\noindent
\textbf{System Prompt to instruct Generator to estimate story point:}
\begin{mdframed}
You are an expert Agile software engineer experienced in story point estimation using the Scrum methodology. Your goal is to estimate story points for new software development issues based on similar past issues retrieved from the same project.\\
You should carefully analyze the reference issues and reason based on their complexity,  to provide a consistent numeric estimate. These numeric estimates should be from the numbers in the fibonacci series (0,1,1,2,3,5,8 and so on) where lower values represent less complex issue and a higher values represent more complex issue.\\
Respond in a clear and concise format.
\end{mdframed}

\noindent
\newline \textbf{User Prompt to instruct Generator to estimate story point:}
\begin{mdframed}
 Below are three similar issues retrieved from the project’s history.
    The issues are ordered by similarity, with the most similar issue listed first.
    Use them as references to estimate the story point for the new issue.\\
    \textbf{\#\#\# Reference Issues:}\\
    {formatted\_similar\_tasks}\\
    \textbf{\#\#\# New Issue to Estimate:}\\
    Title: test\_task\_title\\
    Description: test\_task\_description\\
    \textbf{\#\#\# Instructions:}\\
    1. Compare the new issue with the reference issues.\\
    2. Analyze its relative complexity.\\
    3. Stay within a reasonable numeric range close to the reference story points. Do not make large jumps and do not invent new scales - use the fibonacci series only (0,1,1,2,3,5,8 and so on) and avoid decimal values.\\
    4. Finally, give a single numeric story point value that best fits the new issue, keeping the scale consistent with the reference issues.\\
    \textbf{\#\#\# Output Format}\\
    Estimated Story Point: \texttt{<number>}\\
    Your answer should ONLY CONTAIN YOUR ESTIMATED STORY POINT. DO NOT ELONGATE YOUR ANSWERS.
    
\end{mdframed}
\vspace{6pt}
We followed a trial and error method to obtain the best performing prompt. 
Some of our earlier versions of the prompt didn't include the strict instructions as shown in the User Prompt, because of which, the generator was producing long irrelevant texts instead of estimating the story points. The given example of prompt is used when top\_k=3. We explored with several  temperature settings - 0, 0.1, 0.2, and 0.3. We tested with small values since story point estimation is a prediction on numbers (story points) and by setting the temperature to a low value, we prioritize accuracy over creativity.


%% file: 04_evaluation.tex
\subsection{Evaluation}
Following previous work, we calculated the Mean Absolute Error (MAE) and Median Absolute Error (MdAE) comparing the generated SPs and the ground truth from the test dataset to evaluate our approach. 
The lower the values of MAE and MdAE, the better it is in terms of performance.
MAE and MdAE can be expressed by the following equations.
\begin{equation}
\text{MAE} = \frac{1}{n} \sum_{i=1}^{n} \left| y_{\text{pred}_i} - y_{\text{true}_i} \right|    
\end{equation}

\begin{equation}
\text{MdAE} = \text{median} \left( \left| y_{\text{pred}_i} - y_{\text{true}_i} \right| \right)
\end{equation}

Here, $n$ is the total number of tasks in the test dataset, and $y_{true}$ represents the actual value of the SPs for the tasks. 
These SPs were originally assigned on the projects' Jira project management board and are extracted and stored in Tawosi's dataset. 
On the other hand, $y_{pred}$ is the story point predicted by our generator. 
To measure \textit{MAE}, we calculated the difference between the actual and predicted story points for each of the tasks, and then their overall summation is divided by the total number of tasks in the test dataset. 
\textit{MdAE} is measured by finding out the median of the differences between the predicted and actual story points.
Finally, we compared our results with the results obtained by previous studies using TF-IDF, Deep-SE, LHC-SE, and LHC$_\text{TC}$-SE based approaches on the same dataset.

On the other hand, for the statistical tests to test our hypotheses, we used some statistical testing. One of the tests that we used is the pairwise Wilcoxon signed-rank test. It is a non-parametric statistical test suitable for paired data that does not assume normality. The test evaluates whether the median difference between two paired sets of observations (in this case, the MAE values of two models across the same set of projects) is significantly different from zero. It works by computing the differences between paired observations, ranking the absolute values of these differences, and then comparing the sum of ranks associated with positive differences against those associated with negative differences. A significant result indicates that one method systematically outperforms the other across projects. This test is appropriate for our dataset due to the small sample size and the non-normal distribution of the MAE values. We also used the Kruskal–Wallis test which is a non-parametric alternative to one-way ANOVA and does not assume normality, making it appropriate for our dataset. The test ranks all MAE values from all groups together and evaluates whether the groups come from the same distribution. A significant H statistic value would indicate that at least one approach performs differently from the others.

%% file: 05_results.tex
\section{Results and Discussions}
\label{sec:exp_design_results}
\label{sec:results}
In this section, we describe the findings that we obtained through our experiment. We also discuss how we obtained the results and discuss how we tested our hypotheses. Our results show that the MdAE values are almost the same for most of the projects, but the MAE values differ. Hence to answer our RQs and test our hypotheses, we focus on MAE values in this study.

\subsection{RQ1 - How do the retrieval parameters influence performance, and do the optimal parameter settings differ across project size categories?}
Our first research question explores whether the amount of retrieved context (top\_k) and the randomness in generation (temperature) need to be adjusted depending on how large the project's historical task repository is. 
We computed the mean of MAE for each of the project groups and reported the combination for top\_k and temperature values that gave the lowest mean MAE. Table~\ref{tab:BAAI_parameters} and Table~\ref{tab:SBERT_parameters} show the best top\_k and temperature combination with BAAI and SBERT embedding models respectively. 
The optimal retrieval configuration varied across project sizes and embedding models. 
For the BAAI embedding model, both small and mid-sized projects achieved the lowest MAE with $top\_k = 2$ and $temperature = 0.1$, indicating that retrieving a small number of highly relevant issues and allowing slight generative variability is effective when the project repository is relatively compact. 

In contrast, large projects had the lowest MAE with $top\_k = 4$ and $temperature = 0$, suggesting that larger repositories benefit from retrieving a broader context while maintaining deterministic output. 
For SBERT, however, the optimal settings differed. 
Small projects performed best with $top\_k = 4$, $temperature = 0.1$, while mid-sized projects again favored $top\_k = 2$, $temperature = 0.1$. 
Large projects using SBERT required $top\_k = 4$ and a higher temperature (0.2), indicating the need for both broader retrieval and increased output flexibility. 
These results indicate that project size and embedding choice jointly influence optimal parameter selection, and no single parameter configuration is universally optimal.

\begin{table}[!ht]
\centering
\caption{Best parameter combination across project size groups with BAAI.}
\begin{tabular}{lccc}
\toprule
\textbf{Project Group} & \textbf{Top\_k} & \textbf{Temperature} \\
\midrule
\textbf{Small} & 2 & 0.1\\
\textbf{Mid}   & 2 & 0.1\\
\textbf{Large} & 4 & 0.0 \\
\bottomrule
\end{tabular}
\label{tab:BAAI_parameters}
\end{table}

\begin{table}[h!]
\centering
\caption{Best parameter combination across project size groups with SBERT.}
\begin{tabular}{lccc}
\toprule
\textbf{Project Group} & \textbf{Top\_k} & \textbf{Temperature} \\
\midrule
\textbf{Small} & 4 & 0.1\\
\textbf{Mid}   & 2 & 0.1\\
\textbf{Large} & 4 & 0.2 \\
\bottomrule
\end{tabular}
\label{tab:SBERT_parameters}
\end{table}

\begin{mdframed}
\textit{Answer to RQ1: Retrieval parameters (top\_k and temperature) influence the model performance. However, it does not show a consistent difference across project size groups for RAG with BAAI and SBERT.}
\end{mdframed}






\subsection{RQ2 - Does the effectiveness of the RAG approach for story point estimation differ across small, mid-sized, and large software projects?}

To investigate the effect of project scale, we grouped the 23 projects into small ($\le$500 issues), mid-sized ($\le$2000 issues), and large ($>$2000 issues) categories. 
This grouping was done based on the size clusters in the dataset. 
500 was the threshold point separating the first and the second clusters, 2000 was the threshold point that separates the second and third clusters. 
Two projects were assigned to the category whose overall scale they most closely resembled. For example, the project, \textit{Core Server} (519 tasks) was grouped with small projects, as its size remains substantially closer to other small projects (205–476 tasks) than to the smallest mid-sized project (811 tasks). Similar reasoning was applied to the project ~\textit{DotNetNuke} (2,064 tasks).
Table~\ref{tab:project_sizes} shows the number of tasks in each of the projects categorized according to the project size.

\begin{table}[ht!]
\centering
\caption{Project sizes and task count}
\begin{tabular}{|l|l|c|}
\hline
\textbf{Size} & \textbf{Project Name} & \textbf{\# Tasks} \\
\hline

\multirow{12}{*}{Small} 
& Daemon & 205 \\
& Sawtooth & 206 \\
& Confluence Cloud & 234 \\
& Alloy & 241 \\
& Compass & 260 \\
& CLI & 293 \\
& Fabric & 303 \\
& Duracloud & 310 \\
& Clover & 336 \\
& Confluence Server & 456 \\
& Aptana Studio & 476 \\
& Core Server & 519 \\
\hline

\multirow{6}{*}{Medium}
& Spring XD & 811 \\
& Documentation & 1,005 \\
& Moodle & 1,394 \\
& Sonatype's Nexus & 1,425 \\
& Mesos & 1,513 \\
& DotNetNuke Platform & 2,064 \\
\hline

\multirow{8}{*}{Large}
& Appcelerator Studio & 2,794 \\
& Evergreen & 2,824 \\
& Mule & 2,935 \\
& Titanium & 3,915 \\
& Data Management & 5,381 \\
\hline

\end{tabular}
\label{tab:project_sizes}
\end{table}

\begin{table*}[h!]
\centering
\caption{Performance summary of BAAI across project size groups.}
\begin{tabular}{lccc}
\toprule
\textbf{Project Group} & \textbf{Mean MAE} & \textbf{Standard Deviation} & \textbf{Number of Projects} \\
\midrule
\textbf{Small} & \textbf{1.99} & 1.36 & 12 \\
\textbf{Mid}   & \textbf{1.67} & 0.62 & 6 \\
\textbf{Large} & \textbf{1.90} & 0.75 & 5 \\
\bottomrule
\end{tabular}
\label{tab:BAAI_groups}
\end{table*}

\begin{table*}[h!]
\centering
\caption{Performance summary of SBERT across project size groups.}
\begin{tabular}{lccc}
\toprule
\textbf{Project Group} & \textbf{Mean MAE} & \textbf{Standard Deviation} & \textbf{Number of Projects} \\
\midrule
\textbf{Small} & \textbf{1.90} & 1.26 & 12 \\
\textbf{Mid}   & \textbf{1.61} & 0.57 & 6 \\
\textbf{Large} & \textbf{1.86} & 0.73 & 5 \\
\bottomrule
\end{tabular}
\label{tab:SBERT_groups}
\end{table*}

We used RAG to estimate the story point, computed the Mean Absolute Error (MAE) for each project, and then averaged the values within each size group. 
As shown in Table~\ref{tab:BAAI_groups}, with BAAI embedding model, the mid-sized projects yielded the lowest mean of MAE (1.67), followed by the large (1.90) and small (1.99) groups. 
The same set of information is shown with the SBERT embedding model in Table~\ref{tab:SBERT_groups}. The SBERT embedding model also demonstrates a similar trend with mid-sized projects showing the lowest mean of MAE as 1.61, followed by the large and the small size groups demonstrating the mean of MAE as 1.86 and 1.90 respectively. Small projects exhibited substantially high variance (Standard deviation (SD) = 1.36 with BAAI and SD=1.26 with SBERT), suggesting uneven retrieval effectiveness in smaller repositories. This indicates that for some projects, MAE estimation is very accurate, for example for the project, SERVER, MAE is as low as 0.85 with BAAI embedding model and 0.99 with SBERT while for other projects, for example for the project APSTUD, MAE is as high as 4.98 with BAAI embedding model and 4.66 with SBERT embedding model. On the other hand, for mid-sized projects, SD is as low as 0.62 with BAAI and 0.57 with SBERT embedding model.

However, it is important to see if this difference is meaningful and if the difference is statistically significant. 
For this, we performed the \textit{Kruskal-Wallis} test on both the MAE results with BAAI and SBERT embedding model to see if there is a significant difference in the MAE values. 
Our results showed that there is no statistically significant difference in the performance of our RAG based story point estimation on different project sizes (with BAAI, H = 0.37, p=0.83 and with SBERT, H=0.57, p=0.75). 
For both embedding models, the test did not indicate a significant effect of project size on MAE as $p>0.05$. 
Therefore, we did not find any evidence that the performance of the RAG approach changes with different size of projects. 
Hence, we accept the null hypothesis \textit{${H'}_0$}.\\ 

\begin{mdframed}
\textit{Answer to RQ2: We found no evidence that project size affects RAG-based story point estimation accuracy. \textbf{Therefore, we fail to reject the null hypothesis \textit{${H'}_0$} as stated in subsection~\ref{hypothesis}.}}
\end{mdframed}

\subsection{RQ3 - Does the choice of embedding model significantly affect the estimation accuracy?}
To evaluate whether embedding model choice significantly influenced estimation accuracy, we conducted pairwise \textit{Wilcoxon signed-rank} tests on the MAE values obtained using SBERT and BAAI embedding models for each project size group. 
The results showed no statistically significant differences for small projects ($p = 0.24$), mid-sized projects ($p = 0.44$), or large projects ($p = 0.5$). 
These findings suggest that both SBERT and BAAI embeddings provide comparable retrieval quality within the RAG framework, and the choice of embedding model does not substantially impact the accuracy of story point estimation across different project sizes. To examine overall embedding model sensitivity, we additionally performed a \textit{Wilcoxon signed-rank} test across all 23 projects. 
The results indicated no statistically significant difference between SBERT and BAAI embeddings ($p = 0.16$). 
This supports the conclusion that the choice of embedding model does not substantially impact estimation accuracy in the proposed RAG-based story point estimation framework. Hence, we could not eliminate the null hypothesis \textit{${H''}_0$}.\\

\begin{mdframed}
\textit{Answer to RQ3: There is no significant difference between the accuracy of BAAI and SBERT in predicting the story points for the given projects.\textbf{Therefore, we fail to reject the the null hypothesis \textit{${H''}_0$} as stated in subsection~\ref{hypothesis}.}}
\end{mdframed}

\subsection{RQ4 - How well does our approach perform compared to existing methods?}
Table \ref{tab:final_result} reports the mean absolute errors (MAE) and median absolute errors (MdAE) for all 23 projects across the four baseline methods (Deep-SE, LHC-SE, LHCtc-SE, and TF-IDF) and our two RAG-based models (RAG-SBERT and RAG-BAAI). The MAE and MdAE for our models were computed using the best-performing parameter settings identified in RQ1. The best result for each project is highlighted in red. Based on these results, we counted how often the RAG models outperformed the baselines; these counts are summarized in Table \ref{tab:number_comparison}.

\begin{table*}[ht!]
\small
    \caption{Test Results for LHC-SE, LHC$_\text{TC}$-SE, Deep-SE, TF-IDF, and RAG on Tawosi dataset}
    \centering
    \begin{tabular}{|c|c|c|c|c|c|c|c|c|c|c|}
        \hline
        \multirow{2}{*}{\textbf{Method}} & \multicolumn{2}{c|}{\textbf{Alloy}} & \multicolumn{2}{c|}{\textbf{CLI}} & \multicolumn{2}{c|}{\textbf{Daemon}} & \multicolumn{2}{c|}{\textbf{Clover}} & \multicolumn{2}{c|}{\textbf{Confluence Cloud}} 
        \\
        \cline{2-11}
        & \textbf{MAE} & \textbf{MdAE} & \textbf{MAE} & \textbf{MdAE} & \textbf{MAE} & \textbf{MdAE} & \textbf{MAE} & \textbf{MdAE} & \textbf{MAE} & \textbf{MdAE} 
        \\
        \hline
         LHC-SE & 1.84 & 2.00 & 1.87 & 2.00 & 2.81 & 3.00 & 3.88 & 2.00 & 1.34 & 1.00 \\
         LHC$_\text{TC}$-SE & 2.28 & 2.00 & 1.76 & 2.00 & \textcolor{red}{2.74} & 3.00 & 4.23 & 1.50 & 1.37 & 1.00 \\
         Deep-SE & 1.51 & 1.28 & 1.76 & 1.30 & 3.29 & \textcolor{red}{2.00} & 3.78 & 1.05 & 1.48 & \textcolor{red}{0.93} \\
         TF-IDF & \textcolor{red}{1.44} & 2.00 & 2.98 & 3.00 & \textcolor{red}{2.74} & 3.00 & 4.04 & \textcolor{red}{1.00} & 1.33 & 1.00 \\
         RAG-SBERT & 1.76 & \textcolor{red}{1.00} & \textcolor{red}{1.48} & \textcolor{red}{1.00} & 3.17 & 3.00 & 3.82 & 1.25 & \textcolor{red}{1.30} & 1.00 \\
         RAG-BAAI & 1.90 & \textcolor{red}{1.00} & 1.64 & \textcolor{red}{1.00} & 3.68 & 3.00 & \textcolor{red}{3.66} & 1.50 & 1.55 & 1.00 \\
        \hline
        \multirow{2}{*}{\textbf{Method}} & \multicolumn{2}{c|}{\textbf{Duracloud}} & \multicolumn{2}{c|}{\textbf{Aptana Studio}} & \multicolumn{2}{c|}{\textbf{Confluence Server}} & \multicolumn{2}{c|}{\textbf{Fabric}} & \multicolumn{2}{c|}{\textbf{Sawtooth}} \\
        \cline{2-11}
        & \textbf{MAE} & \textbf{MdAE} & \textbf{MAE} & \textbf{MdAE} & \textbf{MAE} & \textbf{MdAE} & \textbf{MAE} & \textbf{MdAE} & \textbf{MAE} & \textbf{MdAE} \\
        \hline
         LHC-SE & 1.25 & 1.00 & 4.14 & 3.00 & 0.96 & 1.00 & 0.67 & 1.00 & 1.28 & 1.00 \\
         LHC$_\text{TC}$-SE & \textcolor{red}{0.68} & 1.00 & \textcolor{red}{3.99} & 3.00 & 0.96 & 1.00 & \textcolor{red}{0.65} & 1.00 & 0.95 & 1.00 \\
         Deep-SE & \textcolor{red}{0.68} & \textcolor{red}{0.58} & 4.31 & \textcolor{red}{2.70} & \textcolor{red}{0.91} & \textcolor{red}{0.64} & 0.86 & \textcolor{red}{0.71} & 1.18 & 1.12 \\
         TF-IDF & \textcolor{red}{0.68} & 1.00 & \textcolor{red}{3.99} & 3.00 & 0.96 & 1.00 & 1.10 & 1.00 & \textcolor{red}{0.84} & \textcolor{red}{0.00} \\
         RAG-SBERT & 0.79 & 1.00 & 4.66 & 3.00 & 1.07 & 1.00 & 1.24 & 1.00 & 1.1 & 1.00 \\
         RAG-BAAI & 0.79 & 1.00 & 4.98 & 3.00 & 1.15 & 1.00 & 1.28 & 1.00 & 0.88 & \textcolor{red}{0.00} \\

        \hline
    \multirow{2}{*}{\textbf{Method}} & \multicolumn{2}{c|}{\textbf{Compass}} & \multicolumn{2}{c|}{\textbf{Core Server}} & \multicolumn{2}{c|}{\textbf{Spring XD}} & \multicolumn{2}{c|}{\textbf{Documentation}} & \multicolumn{2}{c|}{\textbf{Sonatype's Nexus}} \\
        \cline{2-11}
        & \textbf{MAE} & \textbf{MdAE} & \textbf{MAE} & \textbf{MdAE} & \textbf{MAE} & \textbf{MdAE} & \textbf{MAE} & \textbf{MdAE} & \textbf{MAE} & \textbf{MdAE} \\
        \hline
         LHC-SE & 1.38 & 2.00 & \textcolor{red}{0.85} & 1.00 & 1.54 & \textcolor{red}{1.00} & 2.79 & \textcolor{red}{1.00} & 1.14 & 1.00 \\
         LHC$_\text{TC}$-SE & \textcolor{red}{1.30} & \textcolor{red}{1.00} & \textcolor{red}{0.85} & 1.00 & 1.50 & \textcolor{red}{1.00} & 3.65 & 2.00 & 1.22 & 1.00 \\
         Deep-SE & 1.63 & 1.34 & 0.89 & \textcolor{red}{0.71} & 1.45 & 1.16 & 2.72
 & 1.19 & \textcolor{red}{1.08} & \textcolor{red}{0.88} \\
         TF-IDF & 1.38 & 2.00 & 0.93 & 1.00 & 2.01 & 2.00 & 3.03 & \textcolor{red}{1.00} & 1.17 & 1.00 \\
         RAG-SBERT & 1.38 & \textcolor{red}{1.00} & 0.99 & 1.00 & \textcolor{red}{1.42} & \textcolor{red}{1.00} & \textcolor{red}{2.44} & 2.00 & 1.10 & 1.00 \\
         RAG-BAAI & 1.50
 & \textcolor{red}{1.00} & \textcolor{red}{0.85}
 & 1.00 & 1.66 & \textcolor{red}{1.00} & 2.49 & 2.00 & 1.13 & 1.00 \\

        \hline
    \multirow{2}{*}{\textbf{Method}} & \multicolumn{2}{c|}{\textbf{Moodle}} & \multicolumn{2}{c|}{\textbf{Mule}} & \multicolumn{2}{c|}{\textbf{Appcelerator Studio}} & \multicolumn{2}{c|}{\textbf{Mesos}} & \multicolumn{2}{c|}{\textbf{Titanium}} \\
        \cline{2-11}
        & \textbf{MAE} & \textbf{MdAE} & \textbf{MAE} & \textbf{MdAE} & \textbf{MAE} & \textbf{MdAE} & \textbf{MAE} & \textbf{MdAE} & \textbf{MAE} & \textbf{MdAE} \\
        \hline
         LHC-SE & 6.31 & 7.00 & 2.27 & 2.00 & \textcolor{red}{1.51} & 2.00 & 1.34 & \textcolor{red}{1.00} & 2.53 & 2.00 \\
         LHC$_\text{TC}$-SE & 6.31 & 7.00 & 2.60 & 3.00 & \textcolor{red}{1.51} & 2.00 & \textcolor{red}{1.33} & \textcolor{red}{1.00} & 2.48 & 2.00 \\
         Deep-SE & 3.55 & 2.77
 & 2.24 & \textcolor{red}{1.68}
 & 1.63 & \textcolor{red}{1.38} & 1.34 & 1.12 & \textcolor{red}{2.41} & \textcolor{red}{1.81}
 \\
         TF-IDF & 6.31 & 7.00 & 3.58 & 2.00 & \textcolor{red}{1.51}
 & 2.00 & 1.34 & \textcolor{red}{1.00} & 2.53 & 2.00 \\
         RAG-SBERT & \textcolor{red}{2.14} & \textcolor{red}{2.00} & 2.10 & 2.00 & 2.26 & 2.00 & 1.57 & \textcolor{red}{1.00} & 2.85
 & 2.00 \\
         RAG-BAAI & 2.32 & \textcolor{red}{2.00} & \textcolor{red}{2.03} & 2.00 & 2.14 & 2.00 & 1.48
 & \textcolor{red}{1.00} & 2.87 & 2.00 \\

        \hline
        \multirow{2}{*}{\textbf{Method}} & \multicolumn{2}{c|}{\textbf{DotNetNuke Platform}} & \multicolumn{2}{c|}{\textbf{Data Management}} & \multicolumn{2}{c|}{\textbf{Evergreen}}
        \\
        \cline{2-7}
        & \textbf{MAE} & \textbf{MdAE} & \textbf{MAE} & \textbf{MdAE} & \textbf{MAE} & \textbf{MdAE}  \\
        \cline{1-7}
         LHC-SE & \textcolor{red}{0.71} & 1.00 & 1.56 & 1.00 & \textcolor{red}{0.60} & 1.00 \\
         LHC$_\text{TC}$-SE & \textcolor{red}{0.71} & 1.00 & 1.52 & 1.00 & 0.62 & 1.00 \\
         Deep-SE & 0.72 & \textcolor{red}{0.69} & 1.61 & \textcolor{red}{0.89}
         & 0.63 & \textcolor{red}{0.54} \\
         TF-IDF & 0.79 & 1.00 & \textcolor{red}{1.49} & 1.00 & 0.69 & 1.00 \\
         RAG-SBERT & 1.00 & 1.00 & 1.65 & 1.00 & 0.83 & 1.00 \\
         RAG-BAAI & 0.96 & 1.00 & 1.65 & 1.00 & 0.83
 & 1.00 \\
        \cline{1-7}
    \end{tabular}
    \label{tab:final_result}
\end{table*}

\begin{table*}[ht!]
\small
    \caption{Number of times RAG performed better compared to baseline methods}
    \centering
    \resizebox{\textwidth}{!}{
    \begin{tabular}{|c|c|c|c|c|c|c|c|c|c|c|c|c|}
        \hline
        \multirow{2}{*}{\textbf{Method}} & \multicolumn{4}{c|}{\textbf{Small}} & \multicolumn{4}{c|}{\textbf{Mid-sized}} & \multicolumn{4}{c|}{\textbf{Large}} 
        \\
        \cline{2-13}
        & \textbf{LHC-SE} & \textbf{LHC$_\text{TC}$-SE} & \textbf{Deep-SE} & \textbf{TF-IDF} & \textbf{LHC-SE} & \textbf{LHC$_\text{TC}$-SE} & \textbf{Deep-SE} & \textbf{TF-IDF} & \textbf{LHC-SE} & \textbf{LHC$_\text{TC}$-SE} & \textbf{Deep-SE} & \textbf{TF-IDF} 
        \\
        \hline
         RAG-SBERT & 7 & 5 & 5 & 4 & 3 & 3 & 3 & 4 & 1 & 1 & 1 & 1 \\
         RAG-BAAI & 4 & 4 & 5 & 3 & 3 & 3 & 2 & 4 & 1 & 1 & 1 & 1\\
        \hline
        
    \end{tabular}
    }
    \label{tab:number_comparison}
\end{table*}

\begin{table*}[h!]
\centering
\caption{Wilcoxon Signed-Rank Test Results: RAG-BAAI vs Baseline Methods}
\begin{tabular}{lccc}
\toprule
\textbf{Baseline Method} & \textbf{Small Projects} & \textbf{Medium Projects} & \textbf{Large Projects} \\
\midrule
LHC-SE     & 0.7402 & 0.4219 & 0.9062 \\
LHCtc-SE   & 0.8970 & 0.4219 & 0.8438 \\
Deep-SE    & 0.8833 & 0.5781 & 0.9062 \\
TF-IDF     & 0.9119 & 0.1562 & 0.7812 \\
\bottomrule
\end{tabular}
\label{table:baai_baselines}
\end{table*}

\begin{table*}[h!]
\centering
\caption{Wilcoxon Signed-Rank Test Results: RAG-SBERT vs Baseline Methods}
\begin{tabular}{lccc}
\toprule
\textbf{Baseline Method} & \textbf{Small Projects} & \textbf{Medium Projects} & \textbf{Large Projects} \\
\midrule
LHC-SE     & 0.6499 & 0.2812 & 0.9375 \\
LHCtc-SE   & 0.8494 & 0.2812 & 0.8438 \\
Deep-SE    & 0.6963 & 0.3438 & 0.9375 \\
TF-IDF     & 0.9126 & 0.1562 & 0.7812 \\
\bottomrule
\end{tabular}
\label{table:sbert_baselines}
\end{table*}
RAG-SBERT outperforms LHC-SE on 7 projects and LHCtc-SE on 5 projects. 
Interestingly, although RQ2 showed no performance differences across project-size groups for the RAG models, they still outperform the baselines on several small-sized projects. However, the number of wins alone does not indicate whether the improvements are statistically significant. 
To gain deeper insights and determine whether our models perform significantly better than the baseline models, we conducted pairwise Wilcoxon signed-rank tests within each project-size category.
Table~\ref{table:baai_baselines} and Table~\ref{table:sbert_baselines} shows the Wilcoxon signed-rank test results with the BAAI embedding and the SBERT embedding model respectively. From these two tables we see that across all project size categories, for both embedding variants (SBERT and BAAI), none of the comparisons against the baseline approaches yielded statistically significant differences (all $p>0.05$).
Hence, we could not eliminate the null hypothesis \textit{${H'''}_0$}.

While RAG-based approach does not significantly improve numeric estimation accuracy, it is still comparable to the other discussed baselines. Hence, by retrieving semantically similar historical tasks along with their associated story points, the system can assist developers in reasoning about estimates, particularly in planning meetings.\\ 

\begin{mdframed}
\textit{Answer to RQ4: The MAEs obtained by the RAG approach are not significantly lower than the MAEs of the baseline methods. \textbf{Therefore, we fail to reject the null hypothesis \textit{${H'''}_0$} as stated in subsection~\ref{hypothesis}.} However, RAG-based story point estimation can be beneficial as a decision-support tool where similar historical tasks can be extracted and used by developers during story point estimation.}
\end{mdframed}

%% file: 06_threats.tex
\section{Threats to Validity}
\label{sec:threats_to_validity}

\textbf{\textit{Internal Validity:}} The datasets used in this study have human labeled story points which themselves may have biases. 
Ideally, it is possible to avoid this bias if the dataset contains only those tasks that have been completed satisfying the story points estimated earlier.
Since the dataset was collected from an already published study~\cite{tawosi2022investigating} and it was not possible to identify which tasks were completed satisfying the estimation, the human biases could not be eliminated. 
However, our RAG based approach retrieves multiple similar tasks and allows the model to find the best match to generate the story points.
From the usability perspective, the retrieved similar tasks are exposed to the developers allowing them apply their consent for accepting or rejecting the auto-generated story points.

In the cleaning pipeline we have removed URLs, logs, code blocks.
It is possible that some irrelevant elements may still be there, subjective of the perfection of the cleaning pipeline, which may negatively contribute to the model's performance.
To verify the integrity, we manually checked 100 randomly selected data rows and none of them had any irrelevant elements which gave us confidence on the cleaning pipeline to move ahead.

\textbf{\textit{External Validity:}} Our dataset contains tasks from the open source projects alone which may not reflect the scenarios that are specific to commercial projects. 
However, in this study our dataset contains tasks collected from 23 different projects which provides a good variety of tasks and story point labeling. 

%% file: 07_conclusions.tex
\section{Conclusion \& Future Works}
\label{sec:conclusion}
Story points (SPs) estimate the effort required to complete tasks in Agile development. 
While Planning Poker is the standard approach, it can be time-consuming and subject to human bias. In this study, we introduced a Retrieval-Augmented Generation (RAG) method to automate story point estimation. Our approach retrieves similar historical tasks using embedding-based similarity and feeds them, along with the new task, to a generator model (Llama-3.2-3B Instruct) to produce an estimated SP.

We evaluated performance using Mean Absolute Error and Median Absolute Error, examining how key parameters such as \textit{top\_k} and \textit{temperature} affect accuracy and whether project size moderates these effects. 
After identifying the best-performing parameter configurations, we compared our RAG models against established baselines. Our statistical analysis revealed no significant differences across project size groups (\textit{Kruskal–Wallis} test) and no significant performance differences between the two embedding models (BAAI \textit{bge-large-en-v1.5} and SBERT \textit{all-mpnet-base-v2}). 
While our RAG-based models performed competitively and outperformed certain baselines on individual projects, \textit{Wilcoxon signed-rank} tests showed that they do not consistently surpass all baseline methods. 
Accordingly, we were unable to reject the null hypotheses.
These findings underscore the challenges of applying RAG to structured prediction tasks such as numeric estimation. 
In particular, fully removing human judgment from story point estimation appears to limit performance. 

Future work should explore cleaning or validating task descriptions to reduce noise, as well as constructing datasets collected under controlled, unbiased conditions-ideally in industrial environments. Additional embedding models may also improve retrieval quality.
Since SP estimation relies heavily on historical project context, fine-tuning the generator on project-specific data may enhance accuracy before applying RAG. 
Finally, future studies should consider a broader range of projects, including both open-source and proprietary settings, to better capture the variability of real-world development environments.

%% file: 08_acknowledgement.tex
\section{Acknowledgment}
This research is supported in part by the Natural Sciences and Engineering Research Council of Canada (NSERC) Discovery Grants program, the Canada Foundation for Innovation's John R. Evans Leaders Fund (CFI-JELF), and by the industry-stream NSERC CREATE in Software Analytics Research (SOAR).